\documentclass[fleqn,usenatbib]{mnras}
\usepackage{newtxtext,newtxmath}
\usepackage[T1]{fontenc}
\usepackage{ae,aecompl}
\usepackage{soul}
\DeclareRobustCommand{\VAN}[3]{#2}
\let\VANthebibliography\thebibliography
\def\thebibliography{\DeclareRobustCommand{\VAN}[3]{##3}\VANthebibliography}
\usepackage{graphicx}
\usepackage{amsmath}	
\usepackage{multicol}        
\usepackage{bm}		
\usepackage{pdflscape}	

\title[mnras title]{Probing the Impact of WIMP Dark Matter on Universal Relations, GW170817 Posterior and Radial Oscillations}

\author[Pinku Routaray et al.]{
Pinku Routaray,$^{1}$
Abdul Quddus,$^2$\thanks{E-mail: abdulquddusphy@gmail.com}
Kabir Chakravarti,$^{3,4}$
Bharat Kumar$^{1}$
\\
$^1$Department of Physics and Astronomy, National Institute of Technology, Rourkela, 769008, India\\
$^{2}$Applied Sciences and Humanities Section, University Polytechnic, Aligarh Muslim University, Aligarh, 202002, India\\
$^{3}$Inter-University Centre for Astronomy and Astrophysics, Pune, 411007, India\\
$^{4}$Indian Institute of Technology, Gandhinagar, 382055, India
}

\date{Accepted XXX. Received YYY; in original form ZZZ}

\pubyear{2023}

\begin{document}
\label{firstpage}
\pagerange{\pageref{firstpage}--\pageref{lastpage}}
\maketitle

\begin{abstract}

In this study, we investigate the impact of Weakly Interacting Massive Particles (WIMPs) dark matter (DM) on $C-\Lambda$ universal relations, GW170817 posterior and radial oscillations of neutron stars (NSs) by considering the interactions of uniformly trapped neutralinos as a DM candidate with the hadronic matter through the exchange of the Higgs boson within the framework of the Next-to-Minimal Supersymmetric Standard Model (NMSSM). The hadronic equation of state (EOS) is modeled using the relativistic mean-field (RMF) formalism with IOPB-I, G3, and QMC-RMF series parameter sets. Presence of DM softens the EOS at both the background and the perturbation levels that implies a small shift to the left in the posterior accompanied by a much larger jump in the left of the mass-radius curves with increasing DM mass. It is observed that EOSs with DM also satisfy the $C-\Lambda$ universality relations among their-selves but get slightly shifted to the right in comparison to that without considering DM. Additionally, we find that the inclusion of DM allows the mass-radius ($M-R$) curves to remain consistent with observational constraints for HESS J1731-347, indicating the possibility of classifying it as a dark matter-admixed neutron star (DMANS). Moreover, we explore the impact of DM on the radial oscillations of pulsating stars and investigate the stability of NSs. The results demonstrate a positive correlation between the mass of DM and the frequencies of radial oscillation modes.

\end{abstract}

\begin{keywords}
stars: neutron, (cosmology:) dark matter, (transients:) neutron star mergers, gravitational waves
\end{keywords}

\section{Introduction} 

GW170817, the first discovery of gravitational waves (GWs) from the merger of binary-neutron-stars (BNS) \cite{:2017aa}, has provided a new methodology to constrain an equation-of-state $-$ a definitive relation between pressure and energy density $-$ of NS \citet{Annala:2018aa}. A second signal GW190425, likely emitted during an NS binary coalescence, is originated from a much greater distance \cite{Abbott_2020} and thus, much weaker than GW170817. However, the presence of neutron stars in the binary cannot be established beyond doubt. Nonetheless, possible upcoming detections of such a post-merger signal \citet{Torres-Rivas:2019aa} will put forward yet another macroscopic probe of NS physics, e.g. \cite{Baiotti_2017,Bauswein:2015aa}. The mass constraints put by the GW170817 data ($2.01 \pm 0.04 \lesssim M(M_\odot) \lesssim 2.16 \pm 0.03$) \cite{Rezzolla_2018} and the experimental measured mass of PSR J0348+0432 (with mass $(2.01 \pm 0.04) M_\odot$) \cite{Antoniadis_2013} propound that NSs should have the masses in the limit $\sim 2.0~M_\odot$. The Neutron Star Interior Composition Explorer (NICER) observation constrains the radius of an NS to be in the range ($11.96-14.26$ km) \cite{Miller_2019}. Recently, astronomers made an interesting discovery within the remains of a supernova explosion: a central compact object (CCO) has been seen and named HESS J1731-347 \cite{HESS_2022}. This mysterious object shows a mass of about $0.77^{+0.20}_{-0.17} \ M_\odot$ and a radius estimated to be approximately $10.4^{+0.86}_{-0.78}$ km. The unexpected features of HESS J1731-347 have sparked a lively discussion among scientists, focusing on its true nature. The newly found object's characteristics have raised an intriguing question: is it an NS with an extremely low mass, or could it be a strange star, displaying a more exotic EOS \cite{Horvath-hess_2023,clemente_2023hess, hcdas_2023hess, rather_2023quark,kubis_2023hess,huang_2023hess, sagun_2023hess}? The nature of this compact object is still concealed in mystery, and further observations and thorough studies are needed to uncover its true identity.

The marvelous combination of low temperatures and supranuclear densities, unattended anywhere else in the Universe, makes NSs coveted and peerless laboratories for testing such extreme matter. With the establishment of GW astronomy on a firm observational footing, it is now plausible to comprehend and, better still, probe primal queries of nature. For example, recently in Ref. \cite{Chakravarti_2020}, the effect of introducing extra dimensions on NS properties has been studied, and the brane tension in the Brane-world model is constrained using the GW170817 posterior. Potential forthcoming BNS in-spirals, merger, and post-merger data can be used to test alterations in general relativity with finite-range scalar forces (in particular, $f(r)$ gravity) \cite{Sagunski:2018aa}. The observation of GWs from the BNS merger can also shed light on a plausible model of dark matter (DM) that could yield some extra peaks in the post-merger frequency spectrum. These peaks can be detected in the foreseeable time GWs signal from NS mergers \cite{ELLIS2018607}. Hence, NSs are a sensitive probe to detect and define the nature of DM particles.

Astrophysical constraints on the masses and radii of NSs have long been used to comprehend the microscopic properties of the matter that forms them. Since the first evidence of the "missing mass" problem \cite{zwicky,Zwicky2009}, a lot of theoretical, experimental, and observational efforts have been made to solve the mystery of dark matter. The nature and origin of dark matter particles are still unknown, nevertheless, many dark matter candidates (for a review on dark matter candidates, see e.g. \cite{Taoso_2008}) have been proposed and examined to constrain their properties. An enthralling solution to the dark matter case is given by weakly interacting massive particles (WIMPs) \cite{MU_OZ_2004} (and references therein), which satisfy the relic density of dark matter in the Universe (calculated by the WMAP Collaboration: $0.094 \leq$ relic-density $\leq 0.129$ \cite{Bennett_2003,Spergel_2003}, and the PLANCK Collaboration: 0.120 ± 0.001 \cite{refId0}). However, WIMPs are very arduous to detect, yet their interactions with nuclei through elastic \cite{MU_OZ_2004} or inelastic scattering \cite{ELLIS1988375,EJIRI199314,Starkman:1995aa} are being studied in several laboratories. Experiments for the direct detection of dark matter, such as LUX, XENON, and DAMA/LIBRA, impose stringent constraints on the parameter space of dark matter particles. The stiffest bound on the WIMP-nucleon spin-independent cross-section achieved by the LUX \citet{Silva:2017aa}, XENON \citet{:2017ab}, and DAMA/LIBRA \citet{Belli:2011aa,Bernabei_2019} collaborations are, respectively, $1.1 \times 10^{-46} cm^2$ for a $50 GeV/c^2$ WIMPs, $7.7 \times 10^{-47} cm^2$ for $35 GeV/c^2$ WIMPs, and $\sim 10^{-40} cm^2$ for $10 GeV/c^2$ WIMPs at 90\% C.L. The invisible Higgs decay width, provided by the ATLAS and CMS collaborations at the LHC, strongly constraint the dark matter below $m_h/2$ ($m_h:$ Higgs mass), hence, kinematically favorable for Higgs boson to decay into pair of DM particle with mass $M_{DM} < 62.5 GeV$ \citet{ihbd}. Indirect searches for DM also constrain its mass. The mass of WIMP is constrained in the range of  few eV$<M_{DM}<1$ MeV with the help of the data on the age, density, and observed structure of the present universe \citet{STEIGMAN1985375}. In Ref. \citet{Bringmann:2014aa}, the thermal annihilation rate is constrained with the value $<\sigma v> = 3\times 10^{-26} cm^3s^{-1}$ for DM masses $M_{DM} \leq 100$ GeV ($M_{DM} \leq 200$ GeV) assuming cardinal annihilation to $\mu^+ \mu^-$ ($e^+e^-$) final states and for $M_{DM} \leq 35$ GeV ($M_{DM} \leq 55$ GeV) considering   chief annihilation into light quarks ($b\bar{b}$).  

Apart from DM direct/indirect detections \cite{Gascon_2015,Kahlhoefer_2017,Gaskins_2016}, constraints on DM parameters can also be set by observations of NS properties \cite{Goldman:1989aa,Bertone:2008aa,Kouvaris:2008aa,Lavallaz:2010aa,Kouvaris:2010aa,Kouvaris:2011aa,Leung:2011aa,McDermott:2012aa,Bramante:2013aa,Bramante:2014aa,G_ver_2014,Baryakhtar:2017aa,Raj:2018aa,Bramante:2022aa,ELLIS2018607,Ellis:2018aa,Ivanytskyi:2020aa,Karkevandi:2022aa,Baym:2018aa,McKeen:2018aa,GOULD1990337,Hook2018,Huang:2019aa}. These studies are done by considering WIMPs trapped inside NS \cite{Goldman:1989aa}, studying effects of self-interacting/annihilating DM on NS \citet{Kouvaris:2008aa, Kouvaris:2010aa, Lavallaz:2010aa, Bramante:2013aa, Bramante:2014aa, Ellis:2018aa, Karkevandi:2022aa}, the impact of charged massive DM particle on NSs \cite{GOULD1990337}, probing axions with NS mergers \cite{Hook2018,Huang:2019aa}, and from the collapse of an NS due to accretion of non annihilating DM \cite{Leung:2011aa,Kouvaris:2011aa,McDermott:2012aa,G_ver_2014} etc. In \cite{Nelson_2019,Karkevandi:2022aa}, the DM parameters are constrained by considering DM halo around NS. The effects of DM (WIMPs trapped inside an NS core) on NS properties have been studied \cite{Panotopoulos:2017aa,Das:2019aa,Quddus_2020} within the relativistic mean-field (RMF) formalism. Considering the mechanical model of the dark matter core inside an NS, it is shown in Ref. \cite{ELLIS2018607} that supplementary peaks may be observed in the power spectral density (PSD) of the GW emission following an NS-NS merger. In Ref. \cite{Das:2020aa}, the unknown parameters of the DM EOS are fixed by using Bayesian parameter optimization. Recently, the mass of a Fermionic DM particle ($M_{DM}\sim 60$ GeV for a $2M_\odot$ NS) with its fraction inside an NS is constrained by using data on the spatial distribution of DM in the Milky Way and following the two-fluid approach (DM and NS matter interact gravitationally) \cite{Ivanytskyi:2020aa}. Moreover, in Ref. \cite{Karkevandi:2022aa}, it is found that considering Bosonic DM inside the NS core and imposing constraints from NS observable mass and tidal deformability data by the LIGO/Virgo Collaboration, sub-GeV DM particles are favored with a low fraction below $\sim 5$\%.
In addition, another crucial microscopic property known as the oscillation of an NS plays a significant role in investigating the NS internal composition and serves as a valuable constraint for understanding the EOS. These oscillations can occur in two ways: radial and non-radial oscillations. Over the years, numerous studies have been proposed to elucidate the phenomenon of NS oscillation \citep{Chandrasekhar_1964, Chanmugam_1977, Vaeth_1992, Gondek_1997, Gondek_1999, Kokkotas_2001, athul_2022, souhardya_2023}, each assuming different compositions of the NS. Therefore, in our DMANS model, the impact of dark matter on the oscillations is studied.

In this study, we have considered the presence of WIMPs trapped inside the NS core. The detection of theoretically predicted peaks in the post-merger frequency spectrum \cite{ELLIS2018607} may provide valuable insights into the nature of DM. However, at the present time, the LIGO-VIRGO-KAGRA detectors are still far from achieving the required sensitivity to fully solve the DM puzzle. As discussed, various theoretical studies \cite{ELLIS2018607,Ellis:2018aa,Ivanytskyi:2020aa,Karkevandi:2022aa,Kouvaris:2011aa,Panotopoulos:2017aa,Das:2019aa,Quddus_2020,harishmnras_2020,pinku_prd_2023,pinku_nitr_2023,pinku_hess_2023} lend support to the idea of a DM core existing inside an NS, but the DM parameters exhibit a significant degree of degeneracy. In this work, we present a posterior GW mass-radius contour, developed using data from BNS mergers, to assess which M-R plots (obtained at different WIMP masses) are consistent with the observational data. Additionally, we utilize the mass-radius constraint of the CCO HESS J1731-347 to determine its true nature within the DMANS model. The $C-\Lambda$ universality relations (where $C$ is the compactness of a star and $\Lambda$ is its dimensionless tidal deformability) have been discussed for the EOSs corresponding to these different WIMP masses to investigate whether this universality breaks down with DM presence (and on increasing its mass) or not.

\begin{figure*}
        \includegraphics[width=\columnwidth]{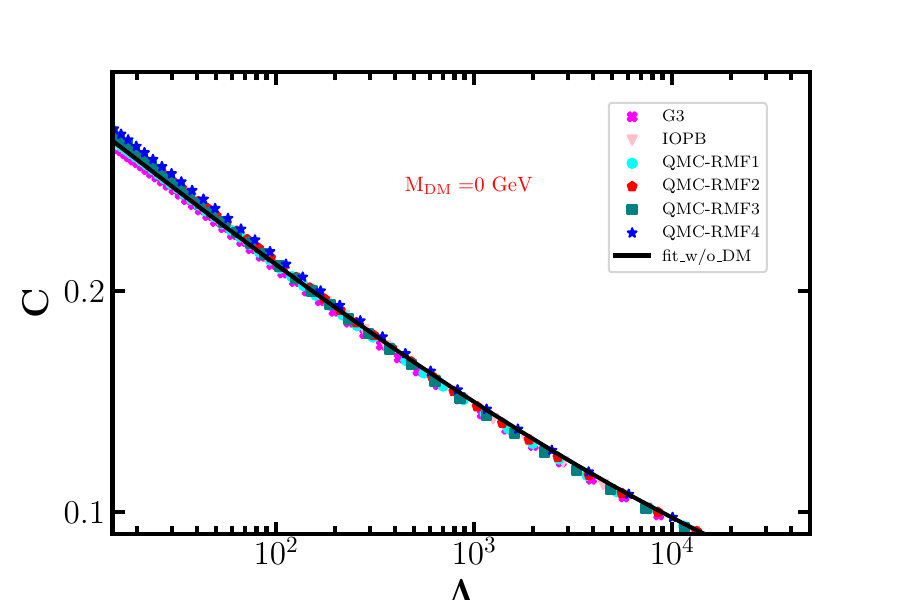}
        \includegraphics[width=\columnwidth]{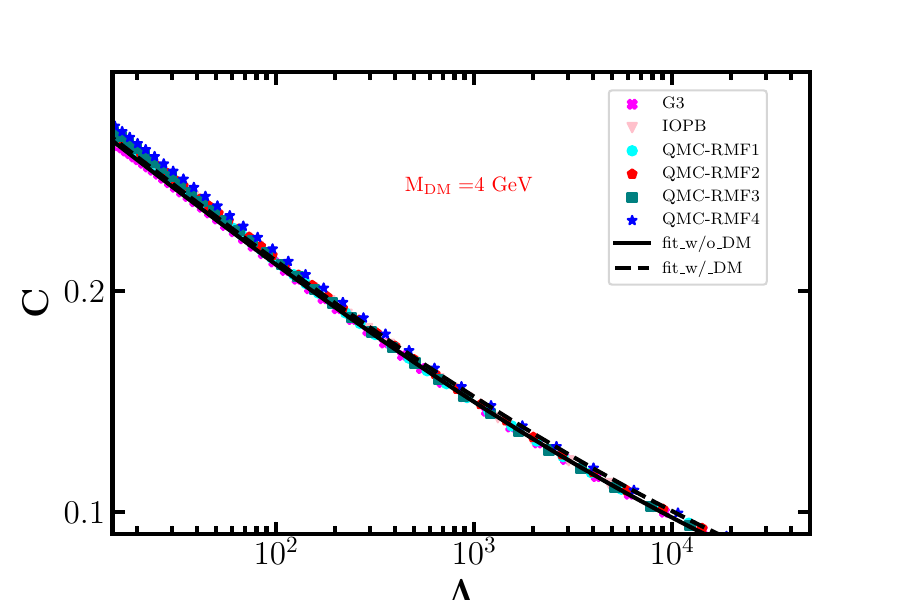} \\
        \includegraphics[width=\columnwidth]{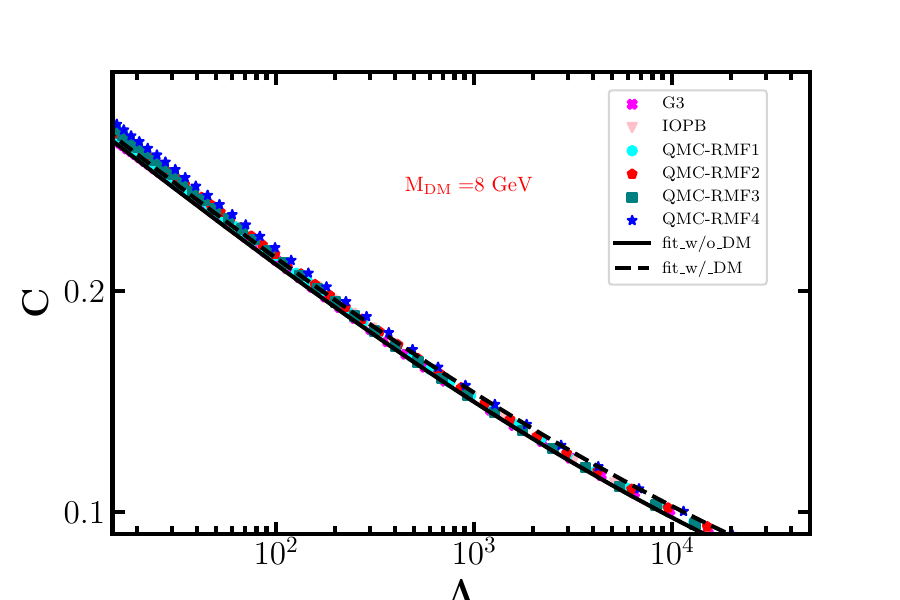}
        \includegraphics[width=\columnwidth]{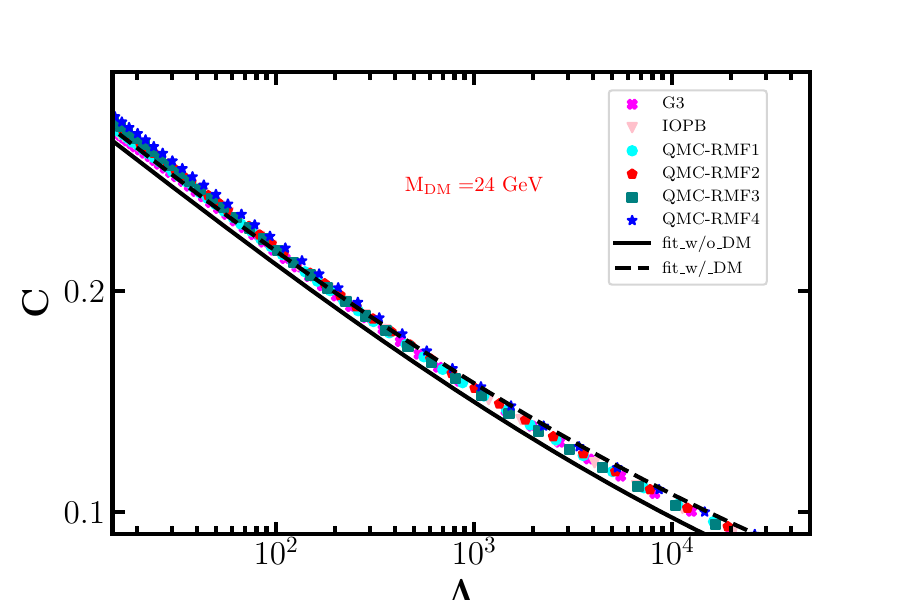}
        \caption{(color online) Universality relation for an NS corresponding to the chosen EOSs without DM is
presented in the panel (a) of the figure, followed by the same in the presence of DM with $M_{DM}= 4$ GeV (b), $M_{DM}= 8$ GeV (c),
and $M_{DM}= 24$ GeV (d), respectively. The black bold line is the fitted curve of $C-\Lambda$ relation when there is no DM presence. The sequence of $C-\Lambda$ points deviate more as more massive DM is added, but does not break the qualitative nature of the universal behavior and the corresponding fitted curves are represented by black dashed lines in Panels (b-d).}
        \label{fig:univ}
\end{figure*}
\medskip

\section{Equation of state of dark matter admixed neutron star}

The presence of a DM core inside an NS with a mass fraction of approximately $\sim 5$\% has significant effects on NS data \cite{Ellis:2018aa}. The amount of DM present inside an NS is not universal; instead, it depends on the environment in which the NS was formed and its evolution history (age, initial temperature) \cite{Ellis:2018aa}. For the usual evolution of an NS with a lifetime of approximately $\sim 10$ Gyr, the accreted WIMP DM amount does not exceed $\sim 10^{-10}M_\odot$ \cite{Goldman:1989aa,Kouvaris:2008aa,Kouvaris:2010aa,G_ver_2014}. The formation of adequate DM cores or halos would require additional mechanisms, such as the conversion of neutrons to scalar DM \cite{Ellis:2018aa}, scalar DM production via the neutron bremsstrahlung process \cite{Ellis:2018aa,Nelson_2019}, or gravitational interaction of the NS with a DM star \cite{Ellis:2018aa, FOOT_2004, FAN2013139, Pollack_2015}. It has been shown in Ref. \cite{Brayeur:2012aa} that BNS mergers might increase the probability of DM agglomeration inside the NS. 

In this study, we have considered the interaction of neutralino ($\chi$) - the lightest mass eigenstate of WIMP within the NMSSM \cite{Gunion:2010aa} - as one of the Fermionic DM candidates. We assume that the neutralino interacts with the hadronic part of the NS through the exchange of the Higgs field $h$. It is important to note that the obtained results and analyses remain valid for any Fermionic WIMP. The reason for considering the WIMP as a DM candidate is that a one-to-one relation is observed between the spin-independent direct detection rate and the DM relic density if the elastic scattering on nuclei occurs dominantly through Higgs exchange \cite{Andreas_2008}.

The NMSSM \cite{ELLWANGER1993331,Ellwanger1995,ELLWANGER199721,MANIATIS_2010} is a simple extension of the MSSM with the addition of a singlet supermultiplet. It offers several advantages over the MSSM: 
\begin{itemize}
\item It provides a solution to the hierarchy problem while preserving the favorable properties of the MSSM.
\item It addresses the $\mu$ problem \cite{KIM1984150}.
\item Compared to the MSSM, NMSSM introduces two additional Higgs bosons.
\end{itemize}
It is noteworthy that in NMSSM \cite{Gunion:2010aa,Bae:2010aa}, the spin-independent cross-section of DM-nucleon scattering can be as high as the one implied by the DAMA results \cite{Andreas_2008}. This is in contrast to the MSSM, where such high cross-sections are not achievable due to the absence of light Higgs bosons exchange. However, it should be mentioned that the DAMA results have faced significant criticism \cite{Buttazzo2020}, and other similar experiments have not confirmed any signal yet \cite{Amar__2020}. In this work, we propose to analyze the effect of this region of the parameter space on the properties of NSs. The interaction Lagrangian density for the neutralino interacting with the hadronic matter of nuclear matter (NM) $(\varphi)$ through Higgs exchange is given as follows.

\begin{eqnarray}
{\cal{L}} & = & {\cal{L}}_{had.} + \bar \chi \left[ i \gamma^\mu \partial_\mu - M_{DM} + y h \right] \chi + 
              \frac{1}{2}\partial_\mu h \partial^\mu h  
\nonumber\\
& &
- \frac{1}{2} M_h^2 h^2 + f \frac{M_n}{v} \bar \varphi h \varphi , 
\label{lag-tot}
\end{eqnarray}
where ${\cal{L}}_{had.}$ is the Lagrangian density for the hadronic matter (for details, see Ref. \cite{Quddus_2020}).
The factor $f$ represents the Higgs-nucleon coupling parameter \cite{Cline:2013aa,Cline:2015aa}, with a central value $f=0.3$ considered based on lattice computations \cite{ALARCON2013413,Young:2013aa,Alvarez-Ruso:2013aa,Ren2012}. In this context, $M_N$ and $M_h$ denote the masses of the nucleon and Higgs boson, respectively, and $v=246~GeV$ corresponds to the Higgs vacuum expectation value. Within NMSSM, a light Higgs boson with mass $M_h=40~GeV$ is chosen to satisfy the condition that the Yukawa coupling (interaction strength of the Higgs field with the DM particles) remains in the perturbative regime ($y < 1$). 
On changing the DM particle mass, the coupling coefficient $y$ is calculated by following the Eq. 
\begin{eqnarray}
\sigma(\chi \varphi \rightarrow \chi \varphi) = \frac{y^2}{\pi}\frac{\mu^2}{v^2 m_h^4}f^2 M_n^2
\end{eqnarray} 
where $\mu$ is the reduced mass of a dark matter-nucleon system, and $\sigma$ is the spin-independent cross-section of dark matter interaction with nucleon, satisfying the DAMA results \cite{Andreas_2008}.

The EOS of NM can be derived within effective theories that incorporate appropriate degrees of freedom. Various approaches, such as non-relativistic Skyrme-type density functional theory \cite{TONDEUR1984297}, perturbative QCD \cite{Kurkela:2010aa,Gorda_2016}, and the relativistic mean-field formalism (RMF) \cite{WALECKA1974491}, are used to generate EOSs of nuclear matter. In this study, we have generated EOSs of pure hadronic matter using the RMF formalism. These EOSs correspond to the parameter sets IOPB-I \cite{Kumar:2018aa}, G3 \cite{KUMAR2017197}, QMC-RMF1, QMC-RMF2, QMC-RMF3, and QMC-RMF4 \cite{PhysRevC.106.055804}. The QMC-RMF series of parameter sets are specifically developed to model nuclear matter across a wide range of temperatures and densities that are relevant for neutron stars and neutron star mergers. These interactions take into account the uncertainties provided by chiral effective field theory ($\chi$EFT) for neutron matter and are consistent with contemporary astrophysical constraints on the EOS \cite{PhysRevC.106.055804}.

By considering the RMF EOS for the hadronic part of the NS and imposing charge neutrality and chemical equilibrium conditions, the EOS of a dark matter-admixed NS can be determined using mean-field approximations and the variational principle. The detailed formalism can be found in Ref. \cite{Quddus_2020}, but for completeness, we provide here the expressions for the energy density and pressure of the dark matter-admixed neutron star.

\begin{eqnarray}
{\cal{E}}_{NS} &=& {\cal{E}}_{had.} + {\cal{E}}_{DM} + {\cal{E}}_{l} \nonumber \\
P_{NS} &=& P_{had.} + P_{DM} + P_{l} .
\label{eos}
\end{eqnarray} 
Where
\begin{eqnarray}
{\cal {E}}_{l} &=& \sum_{l=e,\mu}\frac{2}{(2\pi)^{3}}\int_0^{k_l} d^{3}k \sqrt{k^2 + m_l^2 }, \nonumber \\
P_{l} &=& \sum_{l=e,\mu}\frac{2}{3(2\pi)^{3}}\int_0^{k_l} \frac{d^{3}k k^2} {\sqrt{k^2 + m_l^2}} 
\end{eqnarray}
are the energy density and pressure for the leptonic part of an NS with leptonic mass $m_l$, and fermi momentum $k_l$,
\begin{eqnarray}
{\cal{E}}_{DM} &=& \frac{2}{(2\pi)^{3}}\int_0^{k_F^{DM}} d^{3}k \sqrt{k^2 + (M_{DM}^\star)^2 } + \frac{1}{2}M_h^2 h_0^2 , \nonumber \\
P_{DM} &=& \frac{2}{3(2\pi)^{3}}\int_0^{k_F^{DM}} \frac{d^{3}k k^2} {\sqrt{k^2 + (M_{DM}^\star)^2}} - \frac{1}{2}M_h^2 h_0^2 .
\end{eqnarray}
are the expressions for energy density and pressure for the DM interaction with the hadronic matter of an NS through the exchange of Higgs field with DM fermi momentum $k_f^{DM}=0.06$ GeV \cite{Das:2019aa,Quddus_2020} (assuming baryonic density of NS about $10^3$ times greater than the DM density inside NS). The energy density and pressure for the hadronic part of NS are given as the following:
\begin{eqnarray}
{\cal{E}}_{had.} & = &  \frac{2}{(2\pi)^{3}}\int d^{3}k E_{i}^\ast (k)+\rho_b  W+
\frac{ m_{s}^2\Phi^{2}}{g_{s}^2}\Bigg(\frac{1}{2}+\frac{\kappa_{3}}{3!}
\frac{\Phi }{M} 
\nonumber\\
&&
+ \frac{\kappa_4}{4!}\frac{\Phi^2}{M^2}\Bigg)-\frac{1}{2}m_{\omega}^2\frac{W^{2}}{g_{\omega}^2}\Bigg(1+\eta_{1}\frac{\Phi}{M}+\frac{\eta_2}{2}\frac{\Phi ^2}{M^2}\Bigg)  
\nonumber\\
&&
 -\frac{1}{4!}\frac{\zeta_{0}W^{4}}
        {g_{\omega}^2}+\frac{1}{2}\rho_{3} R
-\frac{1}{2}\Bigg(1+\frac{\eta_{\rho}\Phi}{M}\Bigg)\frac{m_{\rho}^2}{g_{\rho}^2}R^{2} 
\nonumber\\
&&
-\Lambda_{\omega}  (R^{2}\times W^{2})
+\frac{1}{2}\frac{m_{\delta}^2}{g_{\delta}^{2}}\left(D^{2} \right),
\nonumber\\
P_{had.} & = &  \frac{2}{3 (2\pi)^{3}}\int d^{3}k \frac{k^2}{E_{i}^\ast (k)}-
\frac{ m_{s}^2\Phi^{2}}{g_{s}^2}\Bigg(\frac{1}{2}+\frac{\kappa_{3}}{3!}
\frac{\Phi }{M} 
\nonumber\\
& &
+ \frac{\kappa_4}{4!}\frac{\Phi^2}{M^2}  \Bigg)
 +\frac{1}{2}m_{\omega}^2\frac{W^{2}}{g_{\omega}^2}\Bigg(1+\eta_{1}\frac{\Phi}{M}+\frac{\eta_{2}}{2}\frac{\Phi ^2}{M^2}\Bigg) 
 \nonumber\\
& &
 +\frac{1}{4!}\frac{\zeta_{0}W^{4}}{g_{\omega}^2}
+\frac{1}{2}\Bigg(1+\frac{\eta_{\rho}\Phi}{M}\Bigg)\frac{m_{\rho}^2}{g_{\rho}^2}R^{2}
\nonumber\\
& &
+\Lambda_{\omega} (R^{2}\times W^{2})
-\frac{1}{2}\frac{m_{\delta}^2}{g_{\delta}^{2}}\left(D^{2}\right).     
\end{eqnarray}
Where $\Phi$, $D$, $W$, and $R$ are the redefined fields for $\sigma$, $\delta$, $\omega$, and $\rho$ mesons (see \cite{Quddus_2020}), while $E_{i}^\ast(k)$=$\sqrt {k^2+{M_{i}^\ast}^2} \qquad  (i= p,n)$
is the energy with effective mass ${M_{i}^\ast}^2=k_F^2 + M_i^2$, and $k$ is the momentum of the nucleon. The quantities $\rho_b$ and $\rho_3$ in the above equation have their usual meanings as 
the baryonic and iso-scalar densities, respectively. Due to the presence of Higgs field, the effective mass of nucleon get modified as $M_i^\star = M_i + g_\sigma \sigma -\tau_3 g_\delta \delta - \frac{f M_n}{v}h_0$ while the DM effective mass is $M_{DM}^\star = M_{DM} -y h_0$, where $h_0$ is the time component of Higgs field. Putting Eqn. \ref{eos} in the Tolman–Oppenheimer–Volkoff (TOV) equations \citet{Oppenheimer:1939aa,Tolman:1939aa} gives the mass-radius profile of NS.


\begin{figure*}
        \includegraphics[width=\columnwidth]{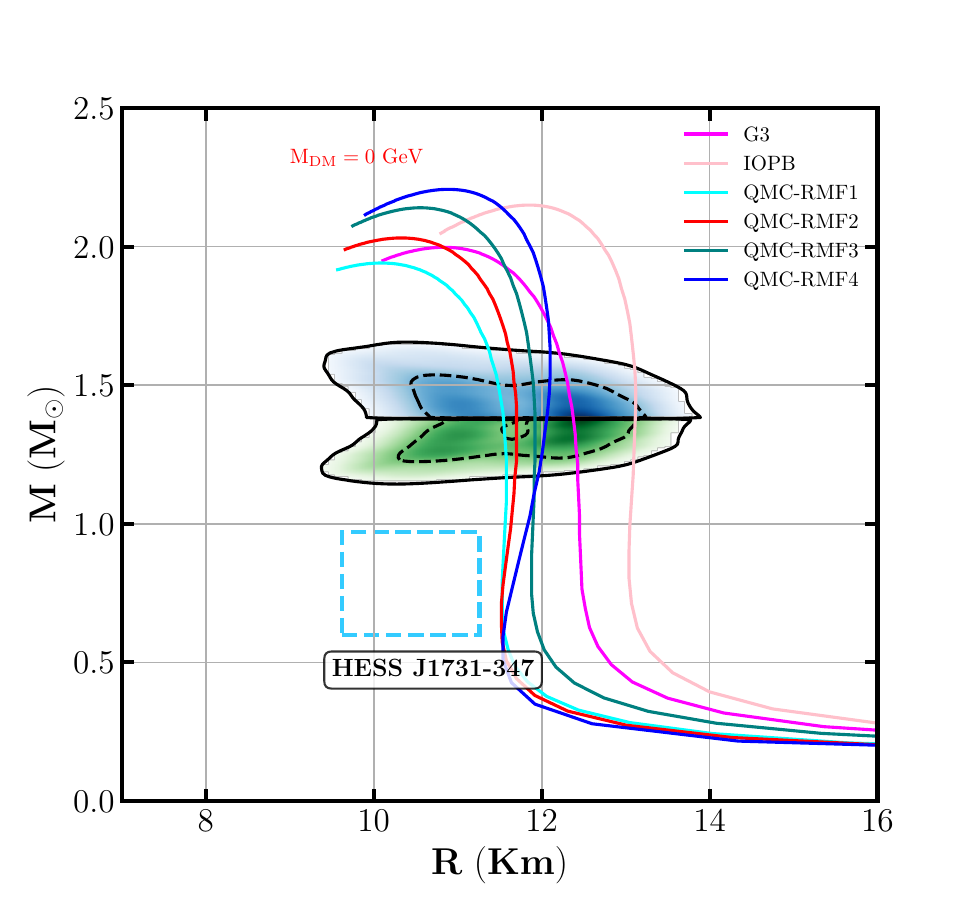}
        \includegraphics[width=\columnwidth]{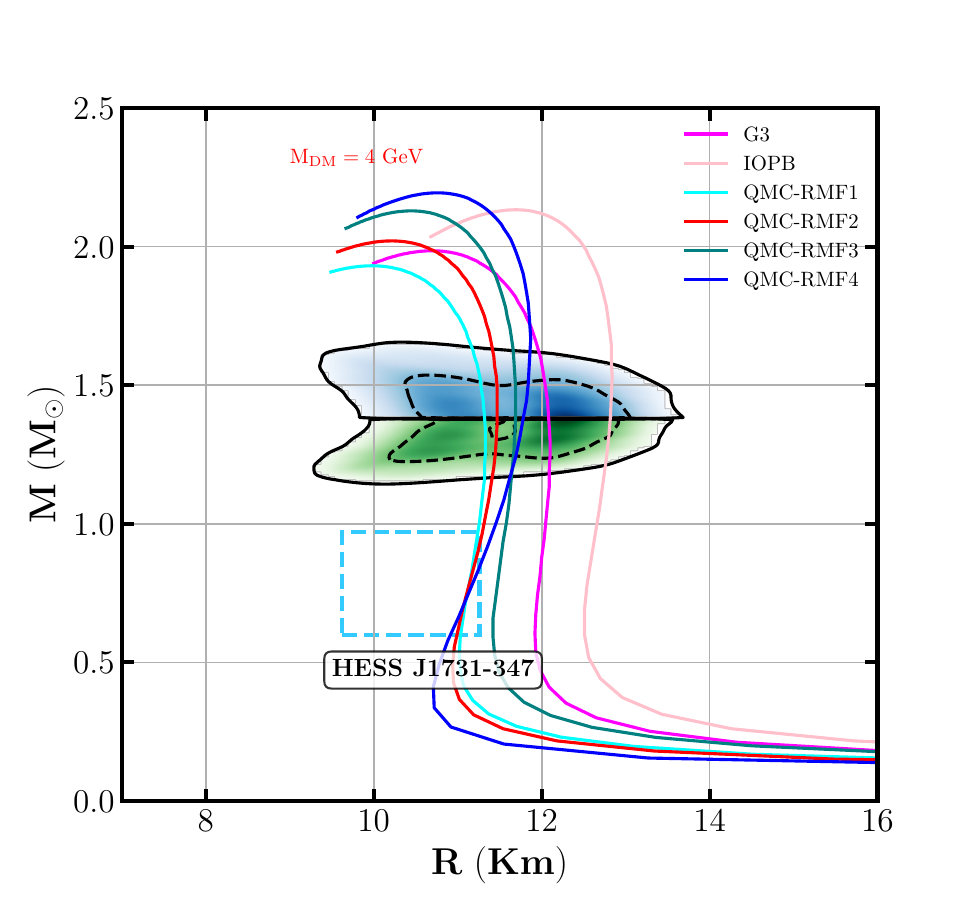} \\
        \includegraphics[width=\columnwidth]{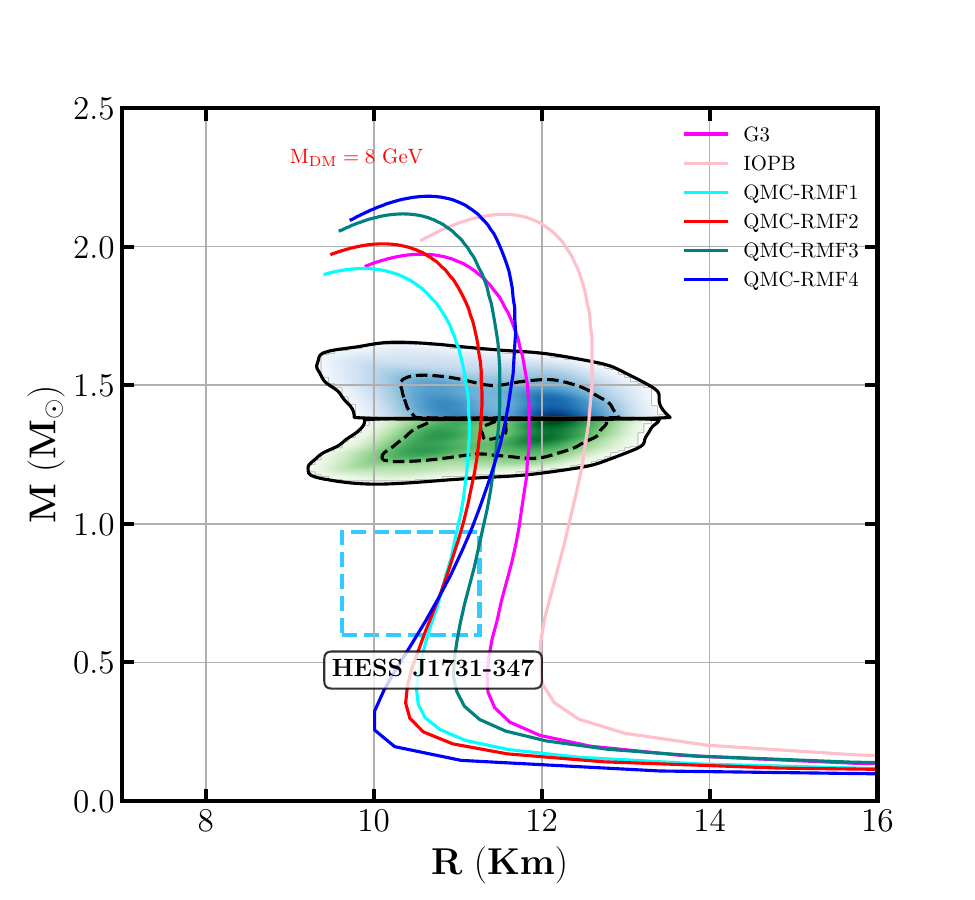}
        \includegraphics[width=\columnwidth]{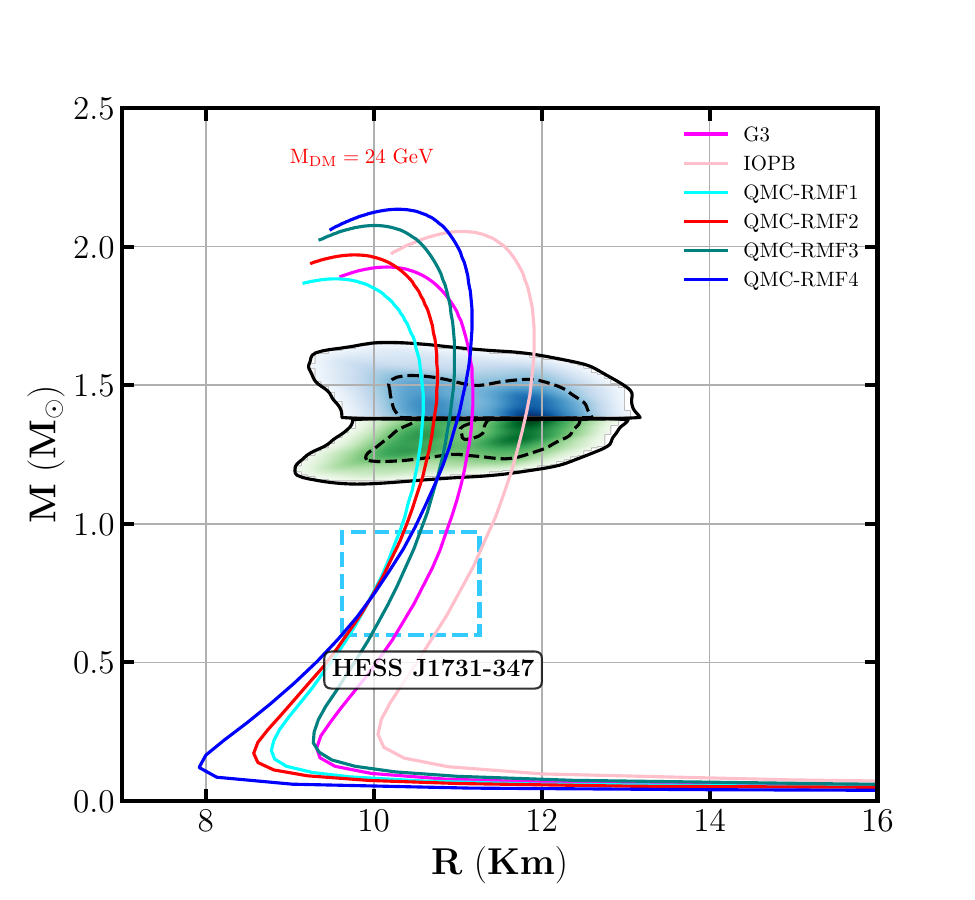}
        \caption{(color online) $M-R$ curves of an NS corresponding to the chosen parameter sets in the absence of DM core (a) and in the presence of DM with $M_{DM}= 4$ GeV (b), $M_{DM}= 8$ GeV (c), and $M_{DM}= 24$ GeV (d). The $90\%$ and $50\%$ C.L. posteriors for the primary (blue) and secondary (green) components of GW170817 are shown bounded by solid lines and dashed lines, respectively, in each panel. The dashed rectangular box shows the observational bound for HESS J1731-347}
        \label{fig:mr}
\end{figure*}

\medskip

\section{Radial oscillations}
To analyze the equations governing the radial oscillations of NS, we introduce the variable $\delta r(r,t)$, representing the time-varying radial displacement of a fluid element. 
\begin{eqnarray}
\delta r(r,t) = X(r)e^{i\omega t}.
\label{eq:standing}
\end{eqnarray}
The variable $X(r)$ represents the amplitude, while $\omega$ denotes the circular frequency of the standing wave solution. The linearized perturbation equation is mathematically represented as shown by \cite{Kokkotas_2001, souhardya_2023},
\begin{equation}\label{eq:radial_equation}
\begin{aligned}
&c_{s}^{2} X^{\prime \prime}+\left(\left(c_{s}^{2}\right)^{\prime}-Z+ \frac{4 \pi G}{c^4} r \gamma P e^{2 \lambda}-\nu^{\prime} c^2 \right) X^{\prime} \\
&+{\left[2\left(\nu^{\prime}\right)^{2}c^2+\frac{2 G m}{r^{3}} e^{2 \lambda}-Z^{\prime}-\frac{4 \pi G }{c^4} (P+\mathcal{E}) Z r e^{2 \lambda}\right]X}\\
&+ \omega^{2} e^{2 \lambda-2 \nu} X =0,
\end{aligned}
\end{equation}
The variable $c_s^2$ represents the square of the speed of sound, while $\gamma$ denotes the adiabatic index. Here, $e^{2\lambda(R)} = \left(1-\frac{2GM}{c^2R}\right)^{-1}$ is the metric function and $Z(r)=\left(\nu^{\prime}-\frac{2}{r}\right)c_{s}^{2}$. And for the numerical solution, we use $\mathcal{E}={\cal{E}}_{NS}$ and $P=P_{NS}$ from equation \ref{eos}.

The boundary conditions for the equations governing oscillations must be formulated to ensure zero displacement at the center ($\delta r(r=0)=0$), while also guaranteeing that the Lagrangian perturbation of pressure becomes zero at the surface ($\Delta P(r=R)=0$).

Considering the mentioned boundary conditions, the displacement function can be rewritten as follow,
\begin{eqnarray}
\zeta = r^2e^{-\nu}X \, .
\label{eq:perturbed_equation}
\end{eqnarray}
By introducing the above new variable, the equation (\ref{eq:radial_equation}) can be reformulated as a Sturm-Liouville differential equation that possesses a self-adjoint characteristic,
\begin{eqnarray}
\frac{d}{dr}\left( H\frac{d \zeta}{d r}\right)+\left(\omega^{2} W+Q\right) \zeta=0 \, ,
\label{eq:self_adjoint_eq}
\end{eqnarray}
where
\begin{align}
&r^2H=(P+\mathcal{E}) e^{\lambda+3 \nu} c_{s}^{2} \, ,
\nonumber \\
&r^2W=(P+\mathcal{E}) e^{3 \lambda+\nu} \, ,
\nonumber \\
&r^2Q=(P+\mathcal{E}) e^{\lambda+3 \nu}\left((\nu^{\prime})^{2}+\frac{4}{r} \nu^{\prime}- \frac{8 \pi G}{c^4} e^{2 \lambda} P\right).
\label{eq:HWQ}
\end{align}
The equation labeled as (\ref{eq:self_adjoint_eq}) represents the fundamental equation governing radial oscillations. In this equation, the quantity $\Delta P$ takes a simple expression given by $\Delta P = -r^{-2} e^{\nu}(P+\mathcal{E}) c_{s}^{2}\zeta^{\prime}$.

Furthermore, it is worth noting that Equation (\ref{eq:self_adjoint_eq}) can be expressed in the Sturm-Liouville form. In this representation, the function $\zeta_n$ exhibits $n$ nodes between the surface and the center and possesses discrete eigenvalues denoted as $\omega_n^2$. These eigenvalues follow a specific structure,
\begin{eqnarray}
    \omega_{0}^{2} < \omega_{1}^{2} <... <  \omega_{n}^{2} < ... .
    \nonumber
    \label{eq:disc}
\end{eqnarray}
The solution represented by Equation (\ref{eq:standing}) indicates that oscillations exhibit harmonic behavior and remain stable when the frequency $\omega$ is real. However, the star will experience instability when the frequency associated with the node is imaginary.

\section{Results and Discussions}
In the literature, it is well known that there exist EOS independent relations \cite{Yagi_2013,Yagi:2013aa,YAGI20171} among NSs. These relations hold true under various circumstances, as long as we exclude cases where abrupt phase transitions occur within the star \cite{YAGI20171}. Of particular interest to us are the $C-\Lambda$ universal relations, also referred to as Universal relations of the First kind.
 This relation can be expressed through the coefficients $c_n$ of the polynomial fit

\begin{equation}\label{eq:fit}
C = \sum_{n=0}^{4} c_n \left( \log_{10}{\Lambda} \right)^n
\end{equation}
between $C$ and $\log_{10}{\Lambda}$. In the presence of DM, both $C$ and $\Lambda$ undergo changes. The panels of Fig.~\ref{fig:univ} display the variation of $C-\Lambda$ across different stellar configurations for four distinct values of DM particle masses. Interestingly, we observe that the sequence of $C-\Lambda$ points deviates more as more massive DM is added, but \textit{does not} break the qualitative nature of the universal behavior. This enables us to fit $C-\Lambda$ for each case of DM mass, and the corresponding fitting coefficients are presented in Table~\ref{tab:fitting_coeff}. 

Since the frequency of gravitational waves depends on the chirp mass and mass ratio, the mass $M$ of a neutron star can be inferred from the observed phases of GWs. Similarly, the tidal deformability $\Lambda$ of the neutron star can also be inferred from the GWs phase during the inspiraling process. Therefore, $(M, \Lambda)$ can be considered as primary GW observables. Subsequently, the $C-\Lambda$ universal relations are utilized to convert the raw $M-\Lambda$ posterior into $M - R$ posteriors \cite{:2017aa}, which are extensively used in EOS inference. In the presence of dark matter, the modified nature of the $C-\Lambda$ function results in a shift in the posterior in the $M-R$ plane, making the EOSs softer, and the $M - R$ curves shift to the left (Fig. \ref{fig:mr}). This shift also affects the posteriors (including the 90\% and the 50\% confidence levels) as shown in Fig. \ref{fig:mr}. A somewhat similar effect was observed in \cite{Chakravarti_2020}, where the presence of extra dimensions resulted in a shift of the $M - R$ curves and posteriors to the right. The effects of changing EOS due to dark matter are not only limited to the background properties of neutron stars but also extend to the perturbations, leading to variations in $C$ for a given value of $\Lambda$ (Fig. \ref{fig:univ}). Consequently, the posteriors shift accordingly in the $M-R$ plane (Fig. \ref{fig:mr}). It is clearly evident that the shifting is more with higher DM mass. Another crucial observation from Fig. \ref{fig:mr} is that the mass-radius profile satisfies the observational bound for HESS J1731-347 when dark matter is considered. Specifically, for M$_{\rm DM}$ = 24 GeV, all the $M-R$ curves satisfy this constraint. As a result, our study also refers the nature of HESS J1731-347 as a dark matter admixed neutron star \cite{sagun_2023hess}.

\begin{table}
	\caption{Fitting coefficients of $C-\Lambda$ relation is shown along with reduced $\chi^2$ value for different DM mass.}
	\label{tab:fitting_coeff}
	\renewcommand{\tabcolsep}{0.08 cm}
	\renewcommand{\arraystretch}{1.0}
	\begin{tabular}{|c|c|c|c|c|c|c|c|c|c|c|c|}
		\hline
		\hline
		M$_{\rm DM}$(GeV) &       $c_0$   &       $c_1$ &$c_2$  &       $c_3$       &       $c_4$       &       $\chi^2_r$(1e-6)        \\
		\hline
		0&  0.34791 & -0.06568 & -0.00358 & 0.00136 & -6.90e-05 & 6.0516\\
		4&  0.35042 & -0.06785 & -0.00232 & 0.00114 & -5.82e-05 & 6.4963\\
		8&  0.35242 & -0.06950 & -0.00140 & 9.97e-04 & -5.07e-05  & 6.8681\\
		24& 0.35821 & -0.07416 & 9.26e-04 & 6.27e-04 & -3.31e-05 & 7.9838\\
		\hline
	\end{tabular} 
\end{table}

\begin{figure}
    \centering
    \includegraphics[width=\columnwidth]{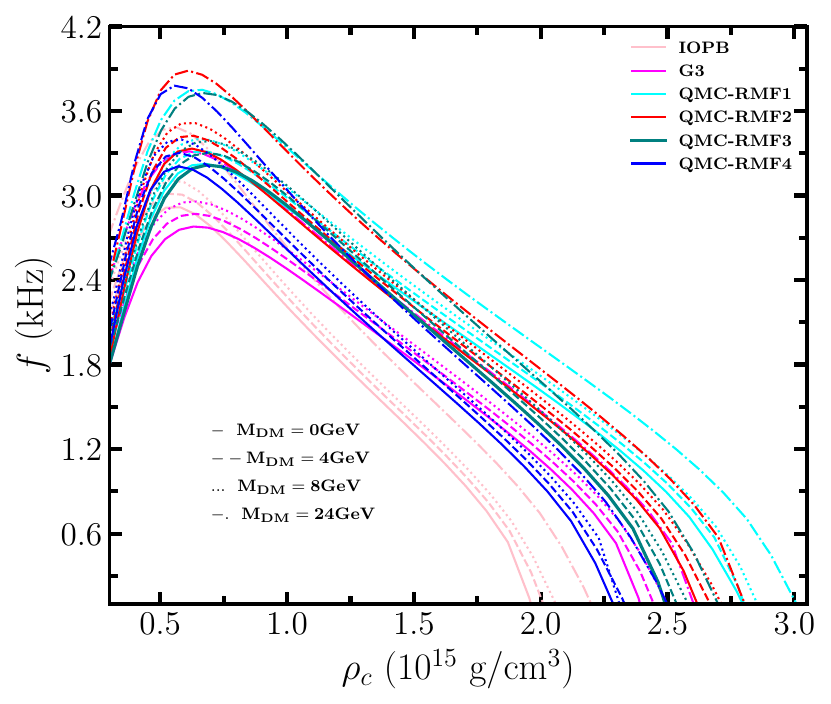}
    \caption{For different DM mass, the radial $f$-mode frequency is shown for G3, IOPB-I, and QMC-RMF series model with varying the central energy density ($\rho_c$).}
    \label{fig:fmode}
\end{figure}

\begin{figure}
    \centering
    \includegraphics[width=\columnwidth]{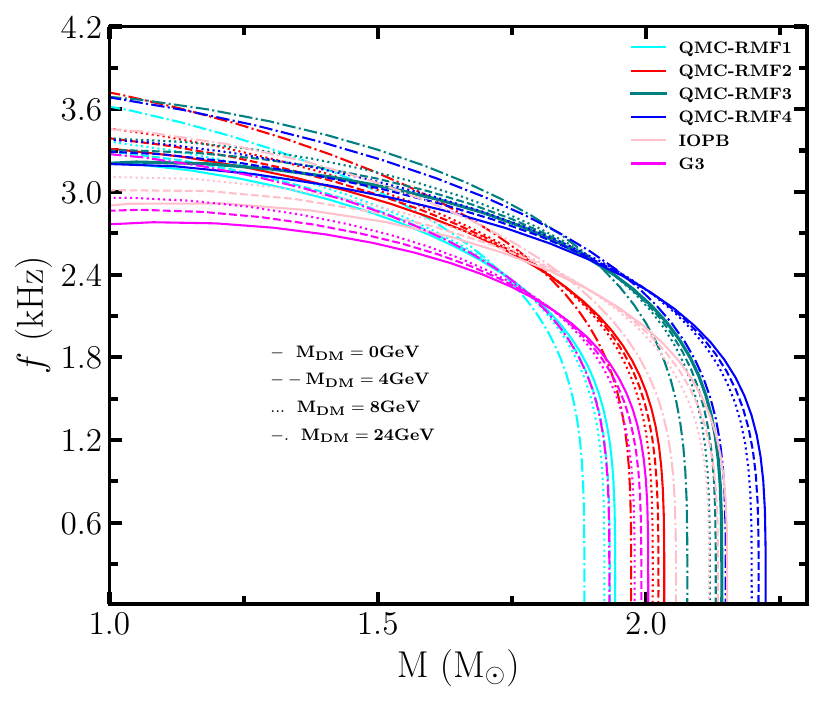}
    \caption{Behaviour of the $f$-mode frequency as a function of mass is plotted for different DM mass.}
    \label{fig:mf}
\end{figure}

In Fig. \ref{fig:fmode}, we investigate the relationship between the eigenfrequencies and central $\rho_c$ for the $f$-mode of radial oscillation. To accomplish this, we vary the mass of the DM using different models. As depicted in the corresponding figure, the stability limit is attained when the density increases and reaches the critical density, regardless of the EOS. At this critical point, the star reaches its maximum mass and becomes unstable, as indicated by a zero eigenvalue for the $f$-mode \citep{Kokkotas_2001}. However, as the mass of DM increases, the critical density also rises due to the softening of the EoS, resulting in higher frequency oscillations. When the density of the star exceeds the critical density, its oscillations exhibit exponential growth. Consequently, the star loses its ability to return to its initial state, ultimately undergoing gravitational collapse and forming a black hole.

Figure \ref{fig:mf} displays how the frequency of the $f$-mode changes with different NS masses, considering the presence and absence of DM, using various models. We observe that the behavior of the $f$-mode rapidly approaches zero precisely at the point where the NS reaches its maximum mass. This finding aligns with the stability criteria $\partial M/\partial\rho_c > 0$, proposed by \cite{Harrison_1965,Shapiro_1983}, which indicate that the NS remains stable as its mass changes with respect to the $\rho_c$.

\section{ Summary and Conclusion}

The primary objective of this work is to investigate the impact of WIMP dark matter on the $C-\Lambda$ universal relation, GW170817 posteriors, and radial $f$-mode oscillation of NSs. We consider interactions of uniformly trapped neutralinos (the lightest WIMP) as a dark matter candidate with the hadronic matter of NSs through the exchange of the Higgs boson, which is the lightest CP-even eigenstate of the NMSSM. The hadronic part of the EOS is modeled using RMF with IOPB-I, G3, and QMC-RMF series of parameter sets. The presence of dark matter inside the NS core results in reduced values of its properties, making the EOSs softer. The findings of this study can be summarized as follows:
\begin{itemize}
	\item The EOSs with dark matter satisfy the universality relation of the first kind, the $C-\Lambda$ universal relations, but they slightly shift to the right with increasing dark matter mass compared to the $C-\Lambda$ universal relation when no dark matter is considered. However, the deviated curves satisfy universality relation among themselves. 
	\item A change in the mass of dark matter results in a small shift to the left in the posterior, accompanied by a much larger jump to the left in the $M-R$ curves.
      \item The incorporation of dark matter into the analysis allows the M-R curve to remain consistent with the mass-radius constraints imposed by HESS J1731-347. Therefore, this study contributes to determining the characteristics of HESS J1731-347, suggesting the possibility that it may be classified as a dark matter-admixed neutron star.
    \item We further investigated the radial oscillations of pulsating stars by solving the Sturm-Liouville equations for the perturbations and imposing necessary boundary conditions. This allowed us to accurately determine the frequencies associated with the mode. The calculations for the fundamental $f-$mode were done with and without the presence of dark matter, while systematically varying the dark matter mass. The numerical findings indicate that the inclusion of dark matter within NSs leads to a reduction in the stiffness of the EOS, resulting in a decrease in the maximum mass of these stars. Additionally, the inclusion of dark matter leads to an increase in the frequencies of pulsating objects. In conclusion, there exists a positive correlation between the mass of the dark matter and the frequencies of the radial oscillation modes.
    \item Finally, we verified the stability criterion of NSs by studying the $f$-mode as a function of their mass.

\end{itemize}


\section*{Acknowledgments}

We are grateful to Sukanta Bose for the discussion and enhancing the content of the paper. A.~Q. is thankful to the IUCAA, Pune for providing nice hospitality during this work. A.~Q. would like to thank S. K. Patra, Shakeb Ahmad, and Grigorios Panotopoulos for the nice discussion. The research of B. K. is supported by DST, India under a SERB CRG Grant with No. CRG/2021/000101.

\section*{Data Availability}

The EOSs used in this article can be inferred from their original articles cited above. Fitting coefficients for $C-\Lambda$ universality relations are given in the text. 

\bibliographystyle{mnras}
\bibliography{posterior} 

\begin{thebibliography}{}
\makeatletter
\relax
\def\mn@urlcharsother{\let\do\@makeother \do\$\do\&\do\#\do\^\do\_\do\%\do\~}
\def\mn@doi{\begingroup\mn@urlcharsother \@ifnextchar [ {\mn@doi@}
  {\mn@doi@[]}}
\def\mn@doi@[#1]#2{\def\@tempa{#1}\ifx\@tempa\@empty \href
  {http://dx.doi.org/#2} {doi:#2}\else \href {http://dx.doi.org/#2} {#1}\fi
  \endgroup}
\def\mn@eprint#1#2{\mn@eprint@#1:#2::\@nil}
\def\mn@eprint@arXiv#1{\href {http://arxiv.org/abs/#1} {{\tt arXiv:#1}}}
\def\mn@eprint@dblp#1{\href {http://dblp.uni-trier.de/rec/bibtex/#1.xml}
  {dblp:#1}}
\def\mn@eprint@#1:#2:#3:#4\@nil{\def\@tempa {#1}\def\@tempb {#2}\def\@tempc
  {#3}\ifx \@tempc \@empty \let \@tempc \@tempb \let \@tempb \@tempa \fi \ifx
  \@tempb \@empty \def\@tempb {arXiv}\fi \@ifundefined
  {mn@eprint@\@tempb}{\@tempb:\@tempc}{\expandafter \expandafter \csname
  mn@eprint@\@tempb\endcsname \expandafter{\@tempc}}}

\bibitem[\protect\citeauthoryear{Abbott et~al.,}{Abbott
  et~al.}{2020}]{Abbott_2020}
Abbott B.~P.,  et~al., 2020, \mn@doi [The Astrophysical Journal Letters]
  {10.3847/2041-8213/ab75f5}, 892, L3

\bibitem[\protect\citeauthoryear{Alarc{\'o}n, {Martin Camalich}  \&
  Oller}{Alarc{\'o}n et~al.}{2013}]{ALARCON2013413}
Alarc{\'o}n J.,  {Martin Camalich} J.,   Oller J.,  2013, \mn@doi [Annals of
  Physics] {https://doi.org/10.1016/j.aop.2013.06.001}, 336, 413

\bibitem[\protect\citeauthoryear{Alford, Brodie, Haber  \& Tews}{Alford
  et~al.}{2022}]{PhysRevC.106.055804}
Alford M.~G.,  Brodie L.,  Haber A.,   Tews I.,  2022, \mn@doi [Phys. Rev. C]
  {10.1103/PhysRevC.106.055804}, 106, 055804

\bibitem[\protect\citeauthoryear{Alvarez-Ruso, Ledwig, Camalich  \&
  Vicente-Vacas}{Alvarez-Ruso et~al.}{2013}]{Alvarez-Ruso:2013aa}
Alvarez-Ruso L.,  Ledwig T.,  Camalich J.~M.,   Vicente-Vacas M.~J.,  2013,
  \mn@doi [Physical Review D] {10.1103/PhysRevD.88.054507}, 88, 054507

\bibitem[\protect\citeauthoryear{Amar{\'{e}} et~al.,}{Amar{\'{e}}
  et~al.}{2020}]{Amar__2020}
Amar{\'{e}} J.,  et~al., 2020, \mn@doi [Journal of Physics: Conference Series]
  {10.1088/1742-6596/1468/1/012014}, 1468, 012014

\bibitem[\protect\citeauthoryear{Andreas, Hambye  \& Tytgat}{Andreas
  et~al.}{2008}]{Andreas_2008}
Andreas S.,  Hambye T.,   Tytgat M. H.~G.,  2008, \mn@doi [Journal of Cosmology
  and Astroparticle Physics] {10.1088/1475-7516/2008/10/034}, 2008, 034

\bibitem[\protect\citeauthoryear{Annala, Gorda, Kurkela  \& Vuorinen}{Annala
  et~al.}{2018}]{Annala:2018aa}
Annala E.,  Gorda T.,  Kurkela A.,   Vuorinen A.,  2018, \mn@doi [Physical
  Review Letters] {10.1103/PhysRevLett.120.172703}, 120, 172703

\bibitem[\protect\citeauthoryear{Antoniadis et~al.,}{Antoniadis
  et~al.}{2013}]{Antoniadis_2013}
Antoniadis J.,  et~al., 2013, \mn@doi [Science] {10.1126/science.1233232}, 340

\bibitem[\protect\citeauthoryear{Baiotti \& Rezzolla}{Baiotti \&
  Rezzolla}{2017}]{Baiotti_2017}
Baiotti L.,  Rezzolla L.,  2017, \mn@doi [Reports on Progress in Physics]
  {10.1088/1361-6633/aa67bb}, 80, 096901

\bibitem[\protect\citeauthoryear{Baryakhtar}{Baryakhtar}{2017}]{Baryakhtar:2017aa}
Baryakhtar M.,  2017, \mn@doi [Physical Review Letters]
  {10.1103/PhysRevLett.119.131801}, 119

\bibitem[\protect\citeauthoryear{Bauswein}{Bauswein}{2015}]{Bauswein:2015aa}
Bauswein A.,  2015, \mn@doi [Physical Review D] {10.1103/PhysRevD.91.124056},
  91

\bibitem[\protect\citeauthoryear{Baym}{Baym}{2018}]{Baym:2018aa}
Baym G.,  2018, \mn@doi [Physical Review Letters]
  {10.1103/PhysRevLett.121.061801}, 121

\bibitem[\protect\citeauthoryear{Belli}{Belli}{2011}]{Belli:2011aa}
Belli P.,  2011, \mn@doi [Physical Review D] {10.1103/PhysRevD.84.055014}, 84

\bibitem[\protect\citeauthoryear{Bennett et~al.,}{Bennett
  et~al.}{2003}]{Bennett_2003}
Bennett C.~L.,  et~al., 2003, \mn@doi [The Astrophysical Journal Supplement
  Series] {10.1086/377253}, 148, 1

\bibitem[\protect\citeauthoryear{Bernabei et~al.,}{Bernabei
  et~al.}{2019}]{Bernabei_2019}
Bernabei R.,  et~al., 2019, \mn@doi [Nuclear Physics and Atomic Energy]
  {10.15407/jnpae2019.04.317}, 20, 317

\bibitem[\protect\citeauthoryear{Bertone}{Bertone}{2008}]{Bertone:2008aa}
Bertone G.,  2008, \mn@doi [Physical Review D] {10.1103/PhysRevD.77.043515}, 77

\bibitem[\protect\citeauthoryear{Bramante}{Bramante}{2013}]{Bramante:2013aa}
Bramante J.,  2013, \mn@doi [Physical Review D] {10.1103/PhysRevD.87.055012},
  87

\bibitem[\protect\citeauthoryear{Bramante}{Bramante}{2014}]{Bramante:2014aa}
Bramante J.,  2014, \mn@doi [Physical Review D] {10.1103/PhysRevD.89.015010},
  89

\bibitem[\protect\citeauthoryear{Bramante}{Bramante}{2022}]{Bramante:2022aa}
Bramante J.,  2022, \mn@doi [Physical Review Letters]
  {10.1103/PhysRevLett.128.231801}, 128

\bibitem[\protect\citeauthoryear{Brayeur}{Brayeur}{2012}]{Brayeur:2012aa}
Brayeur L.,  2012, \mn@doi [Physical Review Letters]
  {10.1103/PhysRevLett.109.061301}, 109

\bibitem[\protect\citeauthoryear{Bringmann}{Bringmann}{2014}]{Bringmann:2014aa}
Bringmann T.,  2014, \mn@doi [Physical Review D] {10.1103/PhysRevD.90.123001},
  90

\bibitem[\protect\citeauthoryear{Buttazzo, Panci, Rossi  \& Strumia}{Buttazzo
  et~al.}{2020}]{Buttazzo2020}
Buttazzo D.,  Panci P.,  Rossi N.,   Strumia A.,  2020, \mn@doi [Journal of
  High Energy Physics] {10.1007/JHEP04(2020)137}, 2020, 137

\bibitem[\protect\citeauthoryear{Chakravarti, Chakraborty, Phukon, Bose  \&
  SenGupta}{Chakravarti et~al.}{2020}]{Chakravarti_2020}
Chakravarti K.,  Chakraborty S.,  Phukon K.~S.,  Bose S.,   SenGupta S.,  2020,
  \mn@doi [Classical and Quantum Gravity] {10.1088/1361-6382/ab8355}, 37,
  105004

\bibitem[\protect\citeauthoryear{{Chandrasekhar}}{{Chandrasekhar}}{1964}]{Chandrasekhar_1964}
{Chandrasekhar} S.,  1964, \mn@doi [apj] {10.1086/147938}, \href
  {https://ui.adsabs.harvard.edu/abs/1964ApJ...140..417C} {140, 417}

\bibitem[\protect\citeauthoryear{{Chanmugam}}{{Chanmugam}}{1977}]{Chanmugam_1977}
{Chanmugam} G.,  1977, \mn@doi [apj] {10.1086/155627}, \href
  {https://ui.adsabs.harvard.edu/abs/1977ApJ...217..799C} {217, 799}

\bibitem[\protect\citeauthoryear{Clemente, Drago  \& Pagliara}{Clemente
  et~al.}{2023}]{clemente_2023hess}
Clemente F.~D.,  Drago A.,   Pagliara G.,  2023, Is the compact object
  associated with HESS J1731-347 a strange quark star? (\mn@eprint {arXiv}
  {2211.07485})

\bibitem[\protect\citeauthoryear{Cline}{Cline}{2013}]{Cline:2013aa}
Cline J.~M.,  2013, \mn@doi [Physical Review D] {10.1103/PhysRevD.88.055025},
  88

\bibitem[\protect\citeauthoryear{Cline}{Cline}{2015}]{Cline:2015aa}
Cline J.~M.,  2015, \mn@doi [Physical Review D] {10.1103/PhysRevD.92.039906},
  92

\bibitem[\protect\citeauthoryear{Das}{Das}{2019}]{Das:2019aa}
Das A.,  2019, \mn@doi [Physical Review D] {10.1103/PhysRevD.99.043016}, 99

\bibitem[\protect\citeauthoryear{Das \& Lopes}{Das \&
  Lopes}{2023}]{hcdas_2023hess}
Das H.~C.,  Lopes L.~L.,  2023, Anisotropic Strange Stars in the Spotlight:
  Unveiling Constraints through Observational Data (\mn@eprint {arXiv}
  {2306.00326})

\bibitem[\protect\citeauthoryear{Das, Malik  \& Nayak}{Das
  et~al.}{2020a}]{Das:2020aa}
Das A.,  Malik T.,   Nayak A.~C.,  2020a, arXiv: 2011.01318

\bibitem[\protect\citeauthoryear{Das, Kumar, Kumar, Biswal, Nakatsukasa, Li  \&
  Patra}{Das et~al.}{2020b}]{harishmnras_2020}
Das H.~C.,  Kumar A.,  Kumar B.,  Biswal S.~K.,  Nakatsukasa T.,  Li A.,
  Patra S.~K.,  2020b, \mn@doi [Monthly Notices of the Royal Astronomical
  Society] {10.1093/mnras/staa1435}, 495, 4893

\bibitem[\protect\citeauthoryear{{Doroshenko}, {Suleimanov}, {P{\"u}hlhofer}
  \& {Santangelo}}{{Doroshenko} et~al.}{2022}]{HESS_2022}
{Doroshenko} V.,  {Suleimanov} V.,  {P{\"u}hlhofer} G.,   {Santangelo} A.,
  2022, \mn@doi [Nature Astronomy] {10.1038/s41550-022-01800-1}, \href
  {https://ui.adsabs.harvard.edu/abs/2022NatAs...6.1444D} {6, 1444}

\bibitem[\protect\citeauthoryear{Ejiri, Fushimi  \& Ohsumi}{Ejiri
  et~al.}{1993}]{EJIRI199314}
Ejiri H.,  Fushimi K.,   Ohsumi H.,  1993, \mn@doi [Physics Letters B]
  {https://doi.org/10.1016/0370-2693(93)91562-2}, 317, 14

\bibitem[\protect\citeauthoryear{Ellis}{Ellis}{2018}]{Ellis:2018aa}
Ellis J.,  2018, \mn@doi [Physical Review D] {10.1103/PhysRevD.97.123007}, 97

\bibitem[\protect\citeauthoryear{Ellis, Flores  \& Lewin}{Ellis
  et~al.}{1988}]{ELLIS1988375}
Ellis J.,  Flores R.,   Lewin J.,  1988, \mn@doi [Physics Letters B]
  {https://doi.org/10.1016/0370-2693(88)91332-9}, 212, 375

\bibitem[\protect\citeauthoryear{Ellis, Hektor, H{\"u}tsi, Kannike, Marzola,
  Raidal  \& Vaskonen}{Ellis et~al.}{2018}]{ELLIS2018607}
Ellis J.,  Hektor A.,  H{\"u}tsi G.,  Kannike K.,  Marzola L.,  Raidal M.,
  Vaskonen V.,  2018, \mn@doi [Physics Letters B]
  {https://doi.org/10.1016/j.physletb.2018.04.048}, 781, 607

\bibitem[\protect\citeauthoryear{Ellwanger, {de Traubenberg}  \&
  Savoy}{Ellwanger et~al.}{1993}]{ELLWANGER1993331}
Ellwanger U.,  {de Traubenberg} M.~R.,   Savoy C.~A.,  1993, \mn@doi [Physics
  Letters B] {https://doi.org/10.1016/0370-2693(93)91621-S}, 315, 331

\bibitem[\protect\citeauthoryear{Ellwanger, Rausch~de Traubenberg  \&
  Savoy}{Ellwanger et~al.}{1995}]{Ellwanger1995}
Ellwanger U.,  Rausch~de Traubenberg M.,   Savoy C.~A.,  1995, \mn@doi
  [Zeitschrift f{\"u}r Physik C Particles and Fields] {10.1007/BF01553993}, 67,
  665

\bibitem[\protect\citeauthoryear{Ellwanger, {Rausch de Traubenberg}  \&
  Savoy}{Ellwanger et~al.}{1997}]{ELLWANGER199721}
Ellwanger U.,  {Rausch de Traubenberg} M.,   Savoy C.,  1997, \mn@doi [Nuclear
  Physics B] {https://doi.org/10.1016/S0550-3213(97)80026-0}, 492, 21

\bibitem[\protect\citeauthoryear{FOOT}{FOOT}{2004}]{FOOT_2004}
FOOT R.,  2004, \mn@doi [International Journal of Modern Physics D]
  {10.1142/s0218271804006449}, 13, 2161

\bibitem[\protect\citeauthoryear{Fan, Katz, Randall  \& Reece}{Fan
  et~al.}{2013}]{FAN2013139}
Fan J.,  Katz A.,  Randall L.,   Reece M.,  2013, \mn@doi [Physics of the Dark
  Universe] {https://doi.org/10.1016/j.dark.2013.07.001}, 2, 139

\bibitem[\protect\citeauthoryear{Gascon}{Gascon}{2015}]{Gascon_2015}
Gascon J.,  2015, \mn@doi [{EPJ} Web of Conferences]
  {10.1051/epjconf/20159502004}, 95, 02004

\bibitem[\protect\citeauthoryear{Gaskins}{Gaskins}{2016}]{Gaskins_2016}
Gaskins J.~M.,  2016, \mn@doi [Contemporary Physics]
  {10.1080/00107514.2016.1175160}, 57, 496

\bibitem[\protect\citeauthoryear{Goldman}{Goldman}{1989}]{Goldman:1989aa}
Goldman I.,  1989, \mn@doi [Physical Review D] {10.1103/PhysRevD.40.3221}, 40,
  3221

\bibitem[\protect\citeauthoryear{{Gondek} \& {Zdunik}}{{Gondek} \&
  {Zdunik}}{1999}]{Gondek_1999}
{Gondek} D.,  {Zdunik} J.~L.,  1999, \mn@doi [aap]
  {10.48550/arXiv.astro-ph/9901167}, \href
  {https://ui.adsabs.harvard.edu/abs/1999A&A...344..117G} {344, 117}

\bibitem[\protect\citeauthoryear{{Gondek}, {Haensel}  \& {Zdunik}}{{Gondek}
  et~al.}{1997}]{Gondek_1997}
{Gondek} D.,  {Haensel} P.,   {Zdunik} J.~L.,  1997, \mn@doi [aap]
  {10.48550/arXiv.astro-ph/9705157}, \href
  {https://ui.adsabs.harvard.edu/abs/1997A&A...325..217G} {325, 217}

\bibitem[\protect\citeauthoryear{Gorda}{Gorda}{2016}]{Gorda_2016}
Gorda T.,  2016, \mn@doi [The Astrophysical Journal]
  {10.3847/0004-637x/832/1/28}, 832, 28

\bibitem[\protect\citeauthoryear{Gould, Draine, Romani  \& Nussinov}{Gould
  et~al.}{1990}]{GOULD1990337}
Gould A.,  Draine B.~T.,  Romani R.~W.,   Nussinov S.,  1990, \mn@doi [Physics
  Letters B] {https://doi.org/10.1016/0370-2693(90)91745-W}, 238, 337

\bibitem[\protect\citeauthoryear{Gunion, V.Belikov  \& Hooper}{Gunion
  et~al.}{2010}]{Gunion:2010aa}
Gunion J.~F.,  V.Belikov A.,   Hooper D.,  2010, arXiv: 1009.2555

\bibitem[\protect\citeauthoryear{G{\"u}ver, Erkoca, Reno  \&
  Sarcevic}{G{\"u}ver et~al.}{2014}]{G_ver_2014}
G{\"u}ver T.,  Erkoca A.~E.,  Reno M.~H.,   Sarcevic I.,  2014, \mn@doi
  [Journal of Cosmology and Astroparticle Physics]
  {10.1088/1475-7516/2014/05/013}, 2014, 013

\bibitem[\protect\citeauthoryear{{Harrison}, {Thorne}, {Wakano}  \&
  {Wheeler}}{{Harrison} et~al.}{1965}]{Harrison_1965}
{Harrison} B.~K.,  {Thorne} K.~S.,  {Wakano} M.,   {Wheeler} J.~A.,  1965,
  {Gravitation Theory and Gravitational Collapse}

\bibitem[\protect\citeauthoryear{Hook \& Huang}{Hook \& Huang}{2018}]{Hook2018}
Hook A.,  Huang J.,  2018, \mn@doi [Journal of High Energy Physics]
  {10.1007/JHEP06(2018)036}, 2018, 36

\bibitem[\protect\citeauthoryear{Horvath, Rocha, de S{\'{a} }, Moraes,
  Bar{\~{a}}o, de Avellar, Bernardo  \& Bachega}{Horvath
  et~al.}{2023}]{Horvath-hess_2023}
Horvath J.~E.,  Rocha L.~S.,  de S{\'{a} } L.~M.,  Moraes P. H. R.~S.,
  Bar{\~{a}}o L.~G.,  de Avellar M. G.~B.,  Bernardo A.,   Bachega R. R.~A.,
  2023, \mn@doi [Astronomy \& Astrophysics] {10.1051/0004-6361/202345885}, 672,
  L11

\bibitem[\protect\citeauthoryear{Huang}{Huang}{2019}]{Huang:2019aa}
Huang J.,  2019, \mn@doi [Physical Review D] {10.1103/PhysRevD.99.063013}, 99

\bibitem[\protect\citeauthoryear{Huang, Hu, Zhang  \& Shen}{Huang
  et~al.}{2023}]{huang_2023hess}
Huang K.,  Hu J.,  Zhang Y.,   Shen H.,  2023, The hadronic equation of state
  of HESS J1731-347 from the relativistic mean-field model with tensor coupling
  (\mn@eprint {arXiv} {2306.04992})

\bibitem[\protect\citeauthoryear{Ivanytskyi}{Ivanytskyi}{2020}]{Ivanytskyi:2020aa}
Ivanytskyi O.,  2020, \mn@doi [Physical Review D]
  {10.1103/PhysRevD.102.063028}, 102

\bibitem[\protect\citeauthoryear{Kahlhoefer}{Kahlhoefer}{2017}]{Kahlhoefer_2017}
Kahlhoefer F.,  2017, \mn@doi [International Journal of Modern Physics A]
  {10.1142/s0217751x1730006x}, 32, 1730006

\bibitem[\protect\citeauthoryear{Karkevandi}{Karkevandi}{2022}]{Karkevandi:2022aa}
Karkevandi D.~R.,  2022, \mn@doi [Physical Review D]
  {10.1103/PhysRevD.105.023001}, 105

\bibitem[\protect\citeauthoryear{Kim \& Nilles}{Kim \&
  Nilles}{1984}]{KIM1984150}
Kim J.~E.,  Nilles H.,  1984, \mn@doi [Physics Letters B]
  {https://doi.org/10.1016/0370-2693(84)91890-2}, 138, 150

\bibitem[\protect\citeauthoryear{Kokkotas \& Ruoff}{Kokkotas \&
  Ruoff}{2001}]{Kokkotas_2001}
Kokkotas K.~D.,  Ruoff J.,  2001, \mn@doi [Astronomy \& Astrophysics]
  {10.1051/0004-6361:20000216}, 366, 565

\bibitem[\protect\citeauthoryear{Kouvaris}{Kouvaris}{2008}]{Kouvaris:2008aa}
Kouvaris C.,  2008, \mn@doi [Physical Review D] {10.1103/PhysRevD.77.023006},
  77

\bibitem[\protect\citeauthoryear{Kouvaris}{Kouvaris}{2010}]{Kouvaris:2010aa}
Kouvaris C.,  2010, \mn@doi [Physical Review D] {10.1103/PhysRevD.82.063531},
  82

\bibitem[\protect\citeauthoryear{Kouvaris}{Kouvaris}{2011}]{Kouvaris:2011aa}
Kouvaris C.,  2011, \mn@doi [Physical Review D] {10.1103/PhysRevD.83.083512},
  83

\bibitem[\protect\citeauthoryear{Kubis, Wójcik, Castillo  \& Zabari}{Kubis
  et~al.}{2023}]{kubis_2023hess}
Kubis S.,  Wójcik W.,  Castillo D.~A.,   Zabari N.,  2023, Relativistic mean
  field model for ultra-compact low mass neutron star of HESS J1731-347
  (\mn@eprint {arXiv} {2307.02979})

\bibitem[\protect\citeauthoryear{Kumar}{Kumar}{2018}]{Kumar:2018aa}
Kumar B.,  2018, \mn@doi [Physical Review C] {10.1103/PhysRevC.97.045806}, 97

\bibitem[\protect\citeauthoryear{Kumar, Singh, Agrawal  \& Patra}{Kumar
  et~al.}{2017}]{KUMAR2017197}
Kumar B.,  Singh S.,  Agrawal B.,   Patra S.,  2017, \mn@doi [Nuclear Physics
  A] {https://doi.org/10.1016/j.nuclphysa.2017.07.001}, 966, 197

\bibitem[\protect\citeauthoryear{Kunjipurayil, Zhao, Kumar, Agrawal  \&
  Prakash}{Kunjipurayil et~al.}{2022}]{athul_2022}
Kunjipurayil A.,  Zhao T.,  Kumar B.,  Agrawal B.~K.,   Prakash M.,  2022,
  \mn@doi [Phys. Rev. D] {10.1103/PhysRevD.106.063005}, 106, 063005

\bibitem[\protect\citeauthoryear{Kurkela}{Kurkela}{2010}]{Kurkela:2010aa}
Kurkela A.,  2010, \mn@doi [Physical Review D] {10.1103/PhysRevD.81.105021}, 81

\bibitem[\protect\citeauthoryear{Kyu Jung~Bae}{Kyu Jung~Bae}{2010}]{Bae:2010aa}
Kyu Jung~Bae H. D.~Kim S.~S.,  2010, \mn@doi [Physical Review D]
  {10.1103/PhysRevD.82.115014}, 82, 115014

\bibitem[\protect\citeauthoryear{Leung}{Leung}{2011}]{Leung:2011aa}
Leung S.-C.,  2011, \mn@doi [Physical Review D] {10.1103/PhysRevD.84.107301},
  84

\bibitem[\protect\citeauthoryear{Lopes}{Lopes}{2022}]{ihbd}
Lopes A.,  2022, ATLAS and CMS chase the invisible with the Higgs boson, \url
  {https://home.cern/news/news/physics/atlas-and-cms-chase-invisible-higgs-boson}

\bibitem[\protect\citeauthoryear{MANIATIS}{MANIATIS}{2010}]{MANIATIS_2010}
MANIATIS M.,  2010, \mn@doi [International Journal of Modern Physics A]
  {10.1142/s0217751x10049827}, 25, 3505

\bibitem[\protect\citeauthoryear{MU{\~{N}}OZ}{MU{\~{N}}OZ}{2004}]{MU_OZ_2004}
MU{\~{N}}OZ C.,  2004, \mn@doi [International Journal of Modern Physics A]
  {10.1142/s0217751x04018154}, 19, 3093

\bibitem[\protect\citeauthoryear{McDermott}{McDermott}{2012}]{McDermott:2012aa}
McDermott S.~D.,  2012, \mn@doi [Physical Review D]
  {10.1103/PhysRevD.85.023519}, 85

\bibitem[\protect\citeauthoryear{McKeen}{McKeen}{2018}]{McKeen:2018aa}
McKeen D.,  2018, \mn@doi [Physical Review Letters]
  {10.1103/PhysRevLett.121.061802}, 121

\bibitem[\protect\citeauthoryear{Miller et~al.,}{Miller
  et~al.}{2019}]{Miller_2019}
Miller M.~C.,  et~al., 2019, \mn@doi [The Astrophysical Journal]
  {10.3847/2041-8213/ab50c5}, 887, L24

\bibitem[\protect\citeauthoryear{Nelson, Reddy  \& Zhou}{Nelson
  et~al.}{2019}]{Nelson_2019}
Nelson A.~E.,  Reddy S.,   Zhou D.,  2019, \mn@doi [Journal of Cosmology and
  Astroparticle Physics] {10.1088/1475-7516/2019/07/012}, 2019, 012

\bibitem[\protect\citeauthoryear{Oppenheimer}{Oppenheimer}{1939}]{Oppenheimer:1939aa}
Oppenheimer J.~R.,  1939, \mn@doi [Physical Review] {10.1103/PhysRev.55.374},
  55, 374

\bibitem[\protect\citeauthoryear{Panotopoulos}{Panotopoulos}{2017}]{Panotopoulos:2017aa}
Panotopoulos G.,  2017, \mn@doi [Physical Review D]
  {10.1103/PhysRevD.96.083004}, 96

\bibitem[\protect\citeauthoryear{{Planck Collaboration} et~al.,}{{Planck
  Collaboration} et~al.}{2020}]{refId0}
{Planck Collaboration} et~al., 2020, \mn@doi [A\&A]
  {10.1051/0004-6361/201833910}, 641, A6

\bibitem[\protect\citeauthoryear{Pollack, Spergel  \& Steinhardt}{Pollack
  et~al.}{2015}]{Pollack_2015}
Pollack J.,  Spergel D.~N.,   Steinhardt P.~J.,  2015, \mn@doi [The
  Astrophysical Journal] {10.1088/0004-637x/804/2/131}, 804, 131

\bibitem[\protect\citeauthoryear{Quddus, Panotopoulos, Kumar, Ahmad  \&
  Patra}{Quddus et~al.}{2020}]{Quddus_2020}
Quddus A.,  Panotopoulos G.,  Kumar B.,  Ahmad S.,   Patra S.~K.,  2020,
  \mn@doi [Journal of Physics G: Nuclear and Particle Physics]
  {10.1088/1361-6471/ab9d36}, 47, 095202

\bibitem[\protect\citeauthoryear{Raj}{Raj}{2018}]{Raj:2018aa}
Raj N.,  2018, \mn@doi [Physical Review D] {10.1103/PhysRevD.97.043006}, 97

\bibitem[\protect\citeauthoryear{Rather, Panotopoulos  \& Lopes}{Rather
  et~al.}{2023}]{rather_2023quark}
Rather I.~A.,  Panotopoulos G.,   Lopes I.,  2023, Quark Models and Radial
  Oscillations: Decoding the HESS J1731-347 Compact Object's Equation of State
  (\mn@eprint {arXiv} {2307.03703})

\bibitem[\protect\citeauthoryear{Ren, Geng, Camalich, Meng  \& Toki}{Ren
  et~al.}{2012}]{Ren2012}
Ren X.-L.,  Geng L.~S.,  Camalich J.~M.,  Meng J.,   Toki H.,  2012, \mn@doi
  [Journal of High Energy Physics] {10.1007/JHEP12(2012)073}, 2012, 73

\bibitem[\protect\citeauthoryear{Rezzolla, Most  \& Weih}{Rezzolla
  et~al.}{2018}]{Rezzolla_2018}
Rezzolla L.,  Most E.~R.,   Weih L.~R.,  2018, \mn@doi [The Astrophysical
  Journal] {10.3847/2041-8213/aaa401}, 852, L25

\bibitem[\protect\citeauthoryear{Routaray, Mohanty, Das, Ghosh, Kalita, Parmar
  \& Kumar}{Routaray et~al.}{2023a}]{pinku_nitr_2023}
Routaray P.,  Mohanty S.~R.,  Das H.~C.,  Ghosh S.,  Kalita P.~J.,  Parmar V.,
   Kumar B.,  2023a (\mn@eprint {arXiv} {2304.05100})

\bibitem[\protect\citeauthoryear{Routaray, Das, Pattnaik  \& Kumar}{Routaray
  et~al.}{2023b}]{pinku_hess_2023}
Routaray P.,  Das H.~C.,  Pattnaik J.~A.,   Kumar B.,  2023b (\mn@eprint
  {arXiv} {2307.12748})

\bibitem[\protect\citeauthoryear{Routaray, Das, Sen, Kumar, Panotopoulos  \&
  Zhao}{Routaray et~al.}{2023c}]{pinku_prd_2023}
Routaray P.,  Das H.~C.,  Sen S.,  Kumar B.,  Panotopoulos G.,   Zhao T.,
  2023c, \mn@doi [Phys. Rev. D] {10.1103/PhysRevD.107.103039}, 107, 103039

\bibitem[\protect\citeauthoryear{Sagun, Giangrandi, Dietrich, Ivanytskyi,
  Negreiros  \& Providência}{Sagun et~al.}{2023}]{sagun_2023hess}
Sagun V.,  Giangrandi E.,  Dietrich T.,  Ivanytskyi O.,  Negreiros R.,
  Providência C.,  2023, What is the nature of the HESS J1731-347 compact
  object? (\mn@eprint {arXiv} {2306.12326})

\bibitem[\protect\citeauthoryear{Sagunski}{Sagunski}{2018}]{Sagunski:2018aa}
Sagunski L.,  2018, \mn@doi [Physical Review D] {10.1103/PhysRevD.97.064016},
  97

\bibitem[\protect\citeauthoryear{Sen, Kumar, Kunjipurayil, Routaray, Ghosh,
  Kalita, Zhao  \& Kumar}{Sen et~al.}{2023}]{souhardya_2023}
Sen S.,  Kumar S.,  Kunjipurayil A.,  Routaray P.,  Ghosh S.,  Kalita P.~J.,
  Zhao T.,   Kumar B.,  2023, \mn@doi [Galaxies] {10.3390/galaxies11020060}, 11

\bibitem[\protect\citeauthoryear{{Shapiro} \& {Teukolsky}}{{Shapiro} \&
  {Teukolsky}}{1983}]{Shapiro_1983}
{Shapiro} S.~L.,  {Teukolsky} S.~A.,  1983, {Black holes, white dwarfs and
  neutron stars. The physics of compact objects},
  \mn@doi{10.1002/9783527617661.
}

\bibitem[\protect\citeauthoryear{Spergel et~al.,}{Spergel
  et~al.}{2003}]{Spergel_2003}
Spergel D.~N.,  et~al., 2003, \mn@doi [The Astrophysical Journal Supplement
  Series] {10.1086/377226}, 148, 175

\bibitem[\protect\citeauthoryear{Starkman}{Starkman}{1995}]{Starkman:1995aa}
Starkman G.~D.,  1995, \mn@doi [Physical Review Letters]
  {10.1103/PhysRevLett.74.2623}, 74, 2623

\bibitem[\protect\citeauthoryear{Steigman \& Turner}{Steigman \&
  Turner}{1985}]{STEIGMAN1985375}
Steigman G.,  Turner M.~S.,  1985, \mn@doi [Nuclear Physics B]
  {https://doi.org/10.1016/0550-3213(85)90537-1}, 253, 375

\bibitem[\protect\citeauthoryear{Taoso, Bertone  \& Masiero}{Taoso
  et~al.}{2008}]{Taoso_2008}
Taoso M.,  Bertone G.,   Masiero A.,  2008, \mn@doi [Journal of Cosmology and
  Astroparticle Physics] {10.1088/1475-7516/2008/03/022}, 2008, 022

\bibitem[\protect\citeauthoryear{Tolman}{Tolman}{1939}]{Tolman:1939aa}
Tolman R.~C.,  1939, \mn@doi [Physical Review] {10.1103/PhysRev.55.364}, 55,
  364

\bibitem[\protect\citeauthoryear{Tondeur, Brack, Farine  \& Pearson}{Tondeur
  et~al.}{1984}]{TONDEUR1984297}
Tondeur F.,  Brack M.,  Farine M.,   Pearson J.,  1984, \mn@doi [Nuclear
  Physics A] {https://doi.org/10.1016/0375-9474(84)90444-5}, 420, 297

\bibitem[\protect\citeauthoryear{Torres-Rivas}{Torres-Rivas}{2019}]{Torres-Rivas:2019aa}
Torres-Rivas A.,  2019, \mn@doi [Physical Review D]
  {10.1103/PhysRevD.99.044014}, 99

\bibitem[\protect\citeauthoryear{{Vaeth} \& {Chanmugam}}{{Vaeth} \&
  {Chanmugam}}{1992}]{Vaeth_1992}
{Vaeth} H.~M.,  {Chanmugam} G.,  1992, aap, \href
  {https://ui.adsabs.harvard.edu/abs/1992A&A...260..250V} {260, 250}

\bibitem[\protect\citeauthoryear{Walecka}{Walecka}{1974}]{WALECKA1974491}
Walecka J.,  1974, \mn@doi [Annals of Physics]
  {https://doi.org/10.1016/0003-4916(74)90208-5}, 83, 491

\bibitem[\protect\citeauthoryear{Yagi}{Yagi}{2013}]{Yagi:2013aa}
Yagi K.,  2013, \mn@doi [Physical Review D] {10.1103/PhysRevD.88.023009}, 88

\bibitem[\protect\citeauthoryear{Yagi \& Yunes}{Yagi \&
  Yunes}{2013}]{Yagi_2013}
Yagi K.,  Yunes N.,  2013, \mn@doi [Science] {10.1126/science.1236462}, 341,
  365

\bibitem[\protect\citeauthoryear{Yagi \& Yunes}{Yagi \&
  Yunes}{2017}]{YAGI20171}
Yagi K.,  Yunes N.,  2017, \mn@doi [Physics Reports]
  {https://doi.org/10.1016/j.physrep.2017.03.002}, 681, 1

\bibitem[\protect\citeauthoryear{Young}{Young}{2013}]{Young:2013aa}
Young R.~D.,  2013, ADP-13-01/T821

\bibitem[\protect\citeauthoryear{Zwicky}{Zwicky}{1933}]{zwicky}
Zwicky F.,  1933, Helv. Phys. Acta, 6, 110

\bibitem[\protect\citeauthoryear{Zwicky}{Zwicky}{2009}]{Zwicky2009}
Zwicky F.,  2009, \mn@doi [General Relativity and Gravitation]
  {10.1007/s10714-008-0707-4}, 41, 207

\bibitem[\protect\citeauthoryear{da Silva}{da~Silva}{2017}]{Silva:2017aa}
da Silva C. F.~P.,  2017, arXiv: 1710.03572

\bibitem[\protect\citeauthoryear{de
  Lavallaz}{de~Lavallaz}{2010}]{Lavallaz:2010aa}
de Lavallaz A.,  2010, \mn@doi [Physical Review D]
  {10.1103/PhysRevD.81.123521}, 81

\bibitem[\protect\citeauthoryear{et al.}{et~al.}{2017a}]{:2017aa}
et al. B.~A.,  2017a, \mn@doi [Physical Review Letters]
  {10.1103/PhysRevLett.119.161101}, 119, 161101

\bibitem[\protect\citeauthoryear{et al.}{et~al.}{2017b}]{:2017ab}
et al. E.~A.,  2017b, \mn@doi [Physical Review Letters]
  {10.1103/PhysRevLett.119.181301}, 119, 181301

\makeatother
\end{thebibliography}





\bsp	
\label{lastpage}
\end{document}